\documentclass{article}
\usepackage{spconf,amsmath,graphicx,hyperref,amssymb}

\usepackage{booktabs}

\title{Evaluating Compositional Structure in Audio Representations}
\name{Chuyang Chen\qquad Bea Steers \qquad Brian McFee \qquad Juan Bello}
\address{Music and Audio Research Laboratory, New York University, USA}
\begin{document}
\ninept
\maketitle
\begin{abstract}
We propose a benchmark for evaluating compositionality in audio representations. Audio compositionality refers to representing sound scenes in terms of constituent sources and attributes, and combining them systematically. While central to auditory perception, this property is largely absent from current evaluation protocols. Our framework adapts ideas from vision and language to audio through two tasks: A-COAT, which tests consistency under additive transformations, and A-TRE, which probes reconstructibility from attribute-level primitives. Both tasks are supported by large synthetic datasets with controlled variation in acoustic attributes, providing the first benchmark of compositional structure in audio embeddings.
\end{abstract}

\begin{keywords}
Audio representation learning, compositionality, audio encoders, evaluation and benchmark
\end{keywords}

\section{Introduction}
\label{sec:intro}

Compositionality---the ability to represent complex structures in terms of their constituent parts and the rules by which they combine---is a hallmark of human perception and cognition~\cite{fodor1988connectionism}. It underpins reasoning, generalization, and ultimately, progress toward general intelligence~\cite{lake2017building}. In the auditory domain, compositionality is especially salient: natural sound scenes emerge as mixtures of individual sources~\cite{bregman1994auditory}, where different acoustic attributes combine in structured but variable ways. 

Despite the inherently compositional nature of sound, modern large-scale pretrained audio encoders are rarely evaluated in this light. Existing benchmarks~\cite{gemmeke2017audio}, \cite{mesaros2017dcase}, \cite{turian2022hear}, \cite{yang2021superb}, \cite{zhang2025icme} primarily focus on downstream classification or recognition tasks. Recent work has proposed evaluation frameworks~\cite{plachouras2025towards} that assess properties including informativeness, equivariance, invariance, and disentanglement. However, it is unclear whether current audio encoders capture compositional structure in their learned embeddings.

To address this gap, we introduce a benchmark for systematically evaluating compositionality in audio representations. Our framework adapts ideas from other domains~\cite{johnson2017clevr}, \cite{andreas2019measuring}, \cite{xie2022coat} to the auditory setting through two complementary tasks, focusing on source- and attribute-level composition. \textbf{A-COAT} (Audio Compositional Object Algebra Test) measures whether encoders preserve global structure under additive transformations of sound mixtures, while \textbf{A-TRE} (Audio Tree Reconstruction Error) provides a graded measure of whether encoders capture local attribute-level composition. Together, these tasks complement recent representation evaluation frameworks by focusing on compositionality as a core property of audio embeddings.

Our contributions are threefold: 
(i) we propose a benchmark\footnote{Code and datasets are available at \url{https://github.com/chuyangchencd/audio-compositionality}.} 
with two complementary tasks—\textbf{A-COAT} and \textbf{A-TRE}—for diagnosing compositional structure in audio representations; 
(ii) we release large-scale and balanced datasets of synthetic audio scenes with controlled variation in acoustic attributes; 
and (iii) we benchmark a diverse set of pretrained audio encoders, establishing reference points for future research on compositionality.

\section{Related Work}
\label{sec:related}

\subsection{Evaluating Audio Representations}

Evaluation of audio representations has largely centered on downstream tasks. Common protocols involve training shallow classification or regression layers on frozen embeddings to assess their utility for tasks such as tagging, speech recognition, or acoustic event detection. Benchmarks including DCASE~\cite{mesaros2017dcase}, HEAR~\cite{turian2022hear}, SUPERB~\cite{yang2021superb}, and the ICME AECC~\cite{zhang2025icme} have expanded this paradigm by aggregating a broad suite of downstream tasks to test robustness and transferability across audio domains. These frameworks provide valuable insight into predictive power and generalization, but they remain tied to task-specific accuracy.

More recent studies have argued for broader evaluation beyond downstream probing~\cite{plachouras2025towards}, emphasizing generalizable and interpretable properties including informativeness, equivariance, invariance, and disentanglement. Although these axes reveal important structural aspects of learned representations, systematic evaluation of compositionality---both in terms of whether embeddings preserve additive mixtures of sources and whether they can be reconstructed from underlying primitive attributes---has not yet been addressed. This leaves open the question of how well current audio encoders capture compositional structure, a capacity central to auditory perception and reasoning.

\subsection{Evaluating Compositionality}

Approaches to compositionality often begin with reasoning tasks.  
In vision and language, CLEVR~\cite{johnson2017clevr}, SCAN~\cite{lake2018generalization}, and related datasets test systematic generalization in question answering or instruction-following setups.  
Audio analogs include CLEAR~\cite{abdelnour2018clear}, which adapts CLEVR to acoustic scenes for QA, and CompA~\cite{ghosh2023compa}, which introduces a diagnostic dataset for compositional reasoning of LLMs. 
Other studies focus on disentangling attributes such as pitch or timbre~\cite{engel2017neural}, or on robustness to perturbations and domain shifts~\cite{wang2023transfer}, \cite{deng2025investigating}, highlighting stability rather than structured composition.  

A complementary line of work directly measures compositionality in the representation space.  
Compositional Object Algebra Test (COAT)~\cite{xie2022coat} tests whether embeddings preserve additive composition under controlled transformations in images, while Tree Reconstruction Error (TRE)~\cite{andreas2019measuring} measures reconstructibility from primitive attributes, applied in both images and text.  
These approaches remain diagnostic tools rather than widely adopted benchmarks.  
We build directly on this line of work, extending COAT and TRE to audio and systematizing them into a benchmark framework for compositionality in audio representations.

\section{Method}
\label{sec:method}

We present the first benchmark explicitly designed to measure compositionality in audio representations.  
Our framework consists of two complementary tasks: \textbf{A-COAT}, which tests whether embeddings preserve algebraic consistency under additive transformations, and \textbf{A-TRE}, which evaluates whether representations can be reconstructed from primitive attributes.  
Together, these perspectives define a unified and reproducible protocol for probing compositional structure in audio embeddings.

\subsection{Audio Scenes}
\label{ssec:scene}
Both proposed tasks are evaluated on \emph{audio scenes}. We define a scene $X$ as a set of $N$ sources $\{s_1, \dots, s_N\}$, where each source is parameterized by four attributes:
(1) timbre, the tone or sound texture of the source;
(2) pitch, the fundamental frequency of the sound;
(3) rate, how quickly the sound repeats over time; and
(4) amplitude, the perceived loudness of the source.
Each attribute is discretized into $K$ classes, and a source is represented as $s_n = [t_n, p_n, r_n, a_n]$.

\subsection{A-COAT}
\label{ssec:coat}

A-COAT tests whether the effect of adding the same sources is represented consistently across different base scenes. 
Each test instance is a quadruple of scenes $(A,B,C,D)$ constructed from a base pair $(A,C)$ and an added set of sources $T=\{s^{+}_1,\dots,s^{+}_t\}$, with
\[
B = A \cup T, \qquad D = C \cup T.
\]

For an encoder $f$ producing embeddings $z_X = f(X)$, we evaluate whether the difference vectors $z_B - z_A$ and $z_D - z_C$ align. 
The A-COAT score is their cosine similarity~\cite{mikolov2013efficient}:
\[
\mathrm{A\text{-}COAT}(A,B,C,D) =
\frac{\langle z_B - z_A,\; z_D - z_C\rangle}
{\|z_B - z_A\|\,\|z_D - z_C\|}.
\]
The score lies in $[-1,1]$, where $1$ indicates perfect alignment, $0$ indicates orthogonality, and $-1$ indicates opposite alignment.

\subsection{A-TRE}
\label{ssec:tre}

A-TRE evaluates whether encoder representations can be systematically constructed from compositional primitives. 
In contrast to A-COAT, which is training-free, A-TRE involves fitting a lightweight neural composition model $g_\theta$ that maps scene metadata to the encoder’s embedding space.  

Let $\mathcal{Y}$ denote the set of all attribute classes across timbre, pitch, rate, and amplitude. 
For each $y \in \mathcal{Y}$ we assign a learnable token vector $Q_y \in \mathbb{R}^D$, where $D$ matches the encoder’s embedding size.  

To compute $g_\theta(X)$ for a scene, each source 
$s_n = [t_n,p_n,r_n,a_n]$ is represented by the sum of the token vectors assigned to its attribute classes: 
$E(s_n) = Q_{t_n} + Q_{p_n} + Q_{r_n} + Q_{a_n}$. 
The sequence $\{E(s_1),\dots,E(s_N)\}$, together with a learnable \texttt{[CLS]} token~\cite{devlin2019bert}, is processed by a single-layer Transformer encoder~\cite{vaswani2017attention} (single-head self-attention + feed-forward), and the output of the \texttt{[CLS]} token defines the predicted scene embedding $\hat z = g_\theta(X)$. This formulation follows TRE in testing linear accessibility of attribute composition, while a scene-level Transformer allows nonlinear aggregation across sources.

The A-TRE score for a scene is the cosine similarity between the encoder and predicted embeddings:
\[
\mathrm{A\text{-}TRE}(X) = \frac{\langle z,\hat z\rangle}{\|z\|\;\|\hat z\|},
\qquad z = f(X).
\]

\begin{table*}[t]
\centering
\renewcommand{\arraystretch}{0.9}
\small
\begin{tabular}{lcccccc}
\toprule
\textbf{Model (checkpoint)} & \textbf{Architecture} & \textbf{Training Objective} & \textbf{\#Params} & \textbf{A-COAT $\uparrow$} & \textbf{A-TRE $\uparrow$} \\
\midrule
Downsample       & –           & None                                & --    & 1.00 $\pm$ 0.01 & 0.23 $\pm$ 0.16 \\
Random           & –           & None                                & --    & 0.00 $\pm$ 0.04 & 0.00 $\pm$ 0.04 \\
\midrule
PANNs (Cnn14)         & CNN         & Supervised classification (AudioSet)   & 81M    & 0.27 $\pm$ 0.24 & 0.93 $\pm$ 0.04 \\
PaSST (PaSST-S)       & Transformer & Supervised classification (AudioSet)   & 86M    & 0.26 $\pm$ 0.19 & 0.87 $\pm$ 0.05 \\
CLAP (630k-AS-best)   & Transformer & Contrastive audio–text pretraining     & 31M    & 0.39 $\pm$ 0.20 & 0.90 $\pm$ 0.05 \\
Whisper (large-v2)    & Hybrid      & Automatic speech recognition (ASR)     & 635M   & 0.32 $\pm$ 0.22 & \underline{0.98 $\pm$ 0.01} \\
AF-Whisper (AF3)      & Hybrid      & Alignment to LLM via adaptor layers & 635M & 0.28 $\pm$ 0.16 & 0.89 $\pm$ 0.03 \\
AudioMAE (AS-2M)      & Transformer & Masked autoencoding (self-supervised)  & 86M    & \textbf{0.41 $\pm$ 0.24} & \textbf{0.99 $\pm$ 0.01} \\
BEATs (iter3)         & Transformer & Iterative masked prediction (self-supervised) & 90M & \underline{0.40 $\pm$ 0.21} & 0.97 $\pm$ 0.02 \\
\bottomrule
\end{tabular}
\caption{Performance comparison across baselines and pretrained audio encoders. The Downsample baseline attains a perfect A-COAT score by construction, while the Random baseline produces near-zero scores on both metrics.}
\label{tab:model_results}
\end{table*}

\section{Experiment}
\label{sec:experiments}

\subsection{Dataset Generation}
\label{ssec:data}

The evaluation of A-TRE requires fine-grained metadata about source attributes, which is difficult to obtain from real recordings. While A-COAT could in principle be applied to real data, we adopt a unified synthetic audio scene generation pipeline to ensure oracle access to attribute primitives and controlled composition structure for both tasks. This design additionally provides precise control over attributes for dataset balancing, as described in Section~\ref{ssec:balance}.

Each scene is a 10-s clip at 32 kHz, composed of $N$ sources. Sources are synthesized with the \texttt{learnfm} DX7 FM synthesizer~\cite{whitman2020learnfm}, controlling timbre and pitch. For each source, we generate a short tone with duration determined by the repetition rate, then duplicate and shift it at regular intervals to fill the 10-s window. Amplitude is applied as a multiplicative gain. The final clip is produced by summing normalized sources, followed by conditional peak normalization to mitigate clipping.

All attributes are discretized into $K=8$ classes:
\begin{itemize}
\item \textbf{Timbre:} eight manually selected \texttt{learnfm} patches.
\item \textbf{Pitch:} MIDI 36–84, linearly binned.
\item \textbf{Rate:} 0.2–3.0 Hz repetition, logarithmically binned.
\item \textbf{Amplitude:} $[-26,0]$ dB, linearly binned and converted to linear gains in $[0,1]$.
\end{itemize}

\textbf{A-COAT:} we generate 50,000 candidate quadruples. For each, a transformation set $T$ of 1–3 sources is sampled, then base scenes $A$ and $C$ are sampled independently with $1$ to $4-|T|$ sources. The completed quadruple is formed as in Section~\ref{ssec:coat}.
\textbf{A-TRE:} we generate 150,000 candidate scenes, each containing 1–4 sources.

\subsection{Data Balancing}
\label{ssec:balance}

To obtain evaluation sets that are both diverse and evenly distributed across attributes,
we apply an entropy-based balancing step to the large candidate pools.
Let $\mathcal{A} = \{\text{timbre}, \text{pitch}, \text{rate}, \text{amplitude}\}$.  
For a scene $X$ and $\alpha \in \mathcal{A}$, with class proportions $p_\alpha(k)$ over $k \in \{1,\dots,K\}$, the normalized \emph{scene-level entropy} for $\alpha$ is
\[
H_\alpha(X) \;=\; -\frac{1}{\log_2 K}\sum_{k=1}^K p_\alpha(k)\log_2 p_\alpha(k),
\]
so that $H_\alpha(X)\in[0,1]$. For an A-COAT test instance $(A,B,C,D)$, we define \emph{quadruple-level entropy} 
for attribute $\alpha \in \mathcal{A}$ as
\[
H_\alpha^{\text{quad}}(A,B,C,D) = H_\alpha(A) + H_\alpha(C) + H_\alpha(T),
\]
where $T$ is the transformation set and $H_\alpha(\cdot)$ is the scene-level attribute entropy.  
This aggregates the entropies of the varying parts, reflecting how both base-scene diversity and transformation diversity contribute to compositional difficulty.

We use \texttt{Entrofy}~\cite{huppenkothen2020entrofy}, a greedy subset selection algorithm that maximizes coverage of user-specified features, to subsample candidate pools.  
Our goal is to obtain approximately uniform distributions of $H_\alpha(X)$ and $H_\alpha^{\text{quad}}(A,B,C,D)$ across $\alpha \in \mathcal{A}$.  
For A-COAT, we subsample 2,000 quadruples from the candidate pool.  
For A-TRE, we subsample 10,000 scenes and partition them into 8,000 training, 1,000 validation, and 1,000 test examples.

For later analysis, we define aggregate scene- and quadruple-level diversity as  
$H(X) = \sum_{\alpha \in \mathcal{A}} H_\alpha(X)$ and  
$H^{\text{quad}}(A,B,C,D) = \sum_{\alpha \in \mathcal{A}} H_\alpha^{\text{quad}}(A,B,C,D)$.  
These summarize overall diversity and are used to analyze how it influences task performance.

\begin{figure}[t]
  \centering
  \includegraphics[width=\columnwidth]{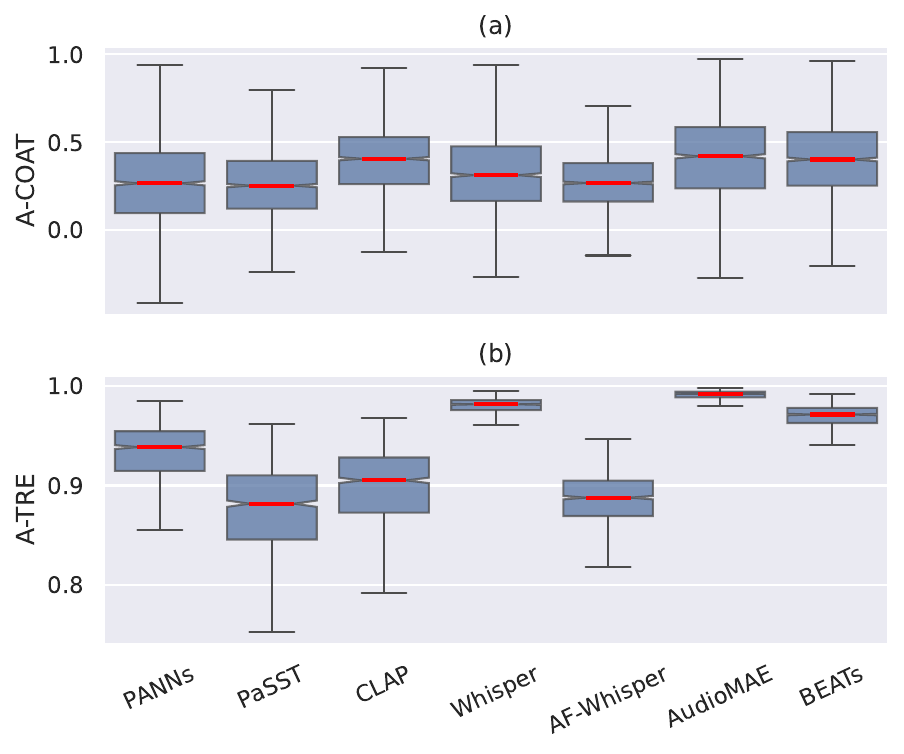}
  \caption{
    Model score distributions for A-COAT (a) and A-TRE (b) as notched box plots. 
    Boxes show the interquartile range with median in red; notches give an approximate 95\% confidence interval.
    }
  \label{fig:dist_coat_tre}
\end{figure}

\begin{figure*}[t]
  \centering
  \includegraphics[width=\textwidth]{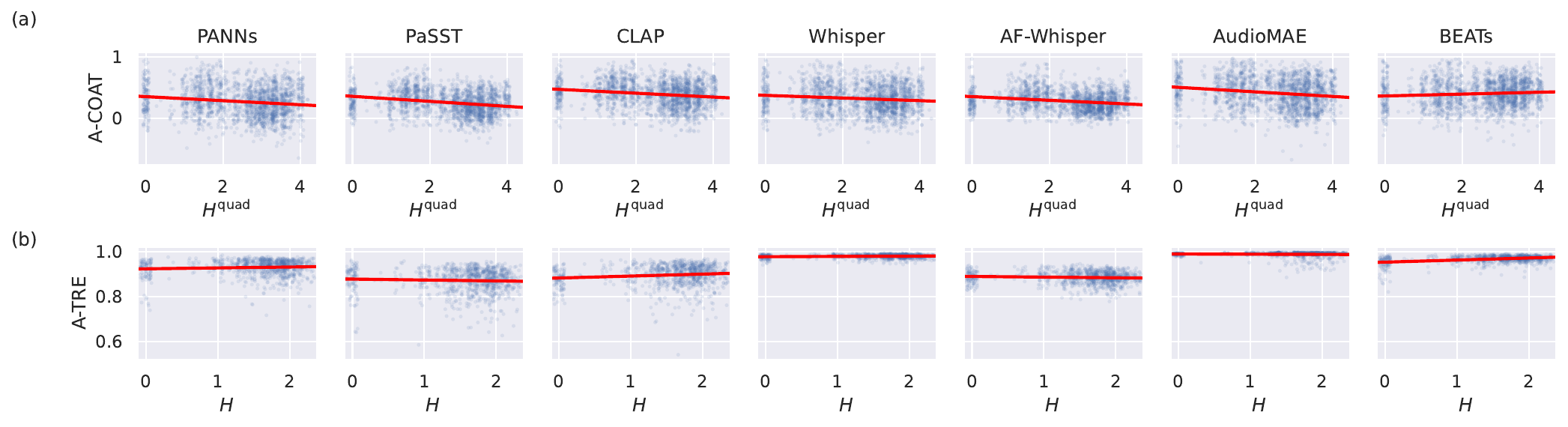}
  \caption{
  Model scores as a function of diversity. 
  (a) A-COAT vs $H^{\mathrm{quad}}$: most models exhibit consistent negative slopes except \textit{BEATs}. 
  (b) A-TRE vs $H$: slopes vary across models, indicating differing sensitivity to diversity. 
  Red lines show linear fits with 95\% confidence intervals. 
  }
  \label{fig:reg_coat_tre}
\end{figure*}

\subsection{Model Selection}
\label{ssec:models}

To contextualize results, we evaluate both simple baselines and a diverse set of pretrained audio encoders, as shown in Table~\ref{tab:model_results}.

\textbf{Baselines.}  
The \textit{Downsample} baseline reduces audio input to dimensionality $D$ via band-limited resampling. Because it preserves linear superposition in the signal domain, it is expected to achieve a trivial perfect score on A-COAT, but to perform poorly on A-TRE as it does not explicitly encode attribute-level composition.
The \textit{Random} baseline samples $D$-dimensional outputs from a normal distribution.
We set $D=768$ for both baselines.

\textbf{Encoders.}  
We include supervised, multimodal, and self-supervised encoders.  
\emph{Supervised:} PANNs~\cite{kong2020panns}, a CNN trained on AudioSet with cross-entropy loss, and PaSST~\cite{koutini2021efficient}, a transformer also trained on AudioSet.  
\emph{Multimodal:} CLAP~\cite{elizalde2023clap}, trained with contrastive audio–text alignment; Whisper~\cite{radford2023robust}, trained for automatic speech recognition (ASR); and AF-Whisper~\cite{goel2025audio}, a Whisper-based encoder adapted for audio question answering.  
\emph{Self-supervised:} AudioMAE~\cite{huang2022masked}, a transformer trained with masked autoencoding, and BEATs~\cite{chen2022beats}, a large-scale SSL model based on iterative pretraining.  
We use officially released pretrained checkpoints for all models.

\subsection{Evaluation Protocol}
\label{ssec:evaluation}

To ensure comparability, we standardize input/output handling: audio is resampled to each model’s required rate; if longer durations are expected, we zero-pad; and if models produce a sequence of embeddings, we global-average to obtain a fixed vector. For Whisper and AF-Whisper, which expect 30-s input, we retain only the first 10-s of tokens before pooling to avoid the influence of padding noise.

For A-COAT, we compute scores directly on the test quadruples using frozen encoder representations; no training is required.  

For A-TRE, evaluation proceeds in two stages. First, for each encoder we train a lightweight composition model $g_\theta$ on the training split with cosine similarity loss and select checkpoints using the validation set. The trained model is then applied to the held-out test set to compute scores.  
TRE models are trained with Adam ($\beta_1{=}0.9$, $\beta_2{=}0.999$), batch size 64, weight decay $10^{-4}$, and learning rate $10^{-4}$ decayed to $10^{-5}$ with cosine annealing, for up to 20 epochs with early stopping after 4 epochs without validation improvement. We do not observe overfitting or underfitting, as confirmed by the small gaps between training and validation curves for all models.

All reported results are averaged across samples for both tasks.

\section{Results and Discussion}
\label{sec:results}

\subsection{Overall Results}
\label{ssec:results}

Table~\ref{tab:model_results} summarizes overall results.  
The baselines behave as expected: \textit{Downsample} achieves a trivial perfect score on A-COAT but fails on A-TRE, underscoring that strong performance on one task does not imply success on the other. \textit{Random} remains near orthogonal on both metrics.  
Across encoders, paired $t$-tests with Benjamini–Hochberg correction~\cite{benjamini1995controlling} indicate significant differences on A-TRE, and most comparisons are also significant on A-COAT.  
The only A-COAT pairs without statistically significant differences are AudioMAE vs.\ BEATs, PANNs vs.\ AF-Whisper, and CLAP vs.\ BEATs.  
These results confirm that both tasks elicit systematic variation across models rather than collapsing to similar scores.
To probe potential effects of the synthetic domain gap, we applied CORAL-based domain adaptation~\cite{sun2016deep} and observed minimal, inconsistent changes, suggesting trends are not driven by simple distributional alignment.

\label{ssec:coat_results}

\subsection{Breakdown: A-COAT Results}

Figure~\ref{fig:dist_coat_tre}(a) shows the distribution of A-COAT scores across models. 
Although overall differences are moderate, models trained with reconstruction-style objectives (AudioMAE, BEATs) and cross-modal alignment (CLAP) consistently reach higher means. 
For AudioMAE and BEATs, this is consistent with their training setup: reconstructing masked patches or acoustic tokens encourages representations that explain mixtures in terms of the additive contributions of individual sources. 
CLAP, despite not being trained on reconstruction, also performs strongly. 
By aligning audio with text embeddings, it inherits a structured space where additive relations are preserved~\cite{mikolov2013efficient}, which supports compositional differences between scenes.  

Figure~\ref{fig:reg_coat_tre}(a) examines how performance varies with quadruple diversity, quantified by $H^{\text{quad}}$. 
All models except BEATs show negative slopes, indicating that their embedding differences become less consistent as $H^{\text{quad}}$ increases. 
This reflects the added challenge of maintaining stable relational structure in more diverse quadruples. 
BEATs is the only model with a positive slope, suggesting that its embeddings become more reliable at higher $H^{\text{quad}}$. 
This distinguishes BEATs from other encoders and shows that models can differ markedly in how they handle higher $H^{\text{quad}}$.

In summary, A-COAT demonstrates that reconstruction-based objectives and cross-modal alignment with representations that themselves encode additive structure (such as text) are especially effective for capturing additive compositional structure. 
Most models are sensitive to increasing diversity, while BEATs uniquely improves, indicating a qualitatively distinct form of robustness.

\subsection{Breakdown: A-TRE Results}
\label{ssec:tre_results}

Figure~\ref{fig:dist_coat_tre}(b) compares model score distributions on A-TRE.  
AudioMAE and BEATs remain strong, consistent with A-COAT, and Whisper also performs well.  
AF-Whisper, though built on the Whisper backbone, is finetuned in the Audio Flamingo framework for audio QA, which prioritizes semantic abstraction over fine-grained acoustic detail.  
This explains its weaker A-TRE performance compared to Whisper.  
PANNs attains relatively strong performance, whereas PaSST—trained with the same tagging objective—performs worse. 
This suggests that CNNs are more effective at capturing the local time–frequency patterns needed for attribute composition than transformer architectures trained for tagging.  
CLAP also ranks lower, as its contrastive audio–text alignment emphasizes invariance rather than detailed attribute encoding.  

Figure~\ref{fig:reg_coat_tre}(b) shows how A-TRE score varies with scene diversity $H$.  
AudioMAE and Whisper remain nearly flat, indicating their embeddings preserve robust attribute composition even as diversity increases.  
The other models display varying patterns, reflecting different ways of generalizing under diversity.  
This underscores that A-TRE exposes diverse generalization behaviors rather than producing a uniform trend across encoders.

In summary, A-TRE demonstrates that local attribute composition is best preserved by models whose objectives emphasize fine-grained acoustic detail (AudioMAE, Whisper, BEATs). In contrast, A-COAT evaluates global relational consistency under additive transformations. Together, the two tasks reveal complementary dimensions of compositionality in audio representations.

\section{Conclusion}
\label{sec:conclusion}

We introduced A-COAT and A-TRE, the first benchmark for systematically evaluating compositionality in audio representations. By adapting ideas from vision and language to the auditory domain, our framework provides complementary measures of global relational structure and local attribute-level composition.

Experiments across pretrained audio encoders reveal clear differences in how training paradigms affect compositional structure, with self-supervised models such as AudioMAE and BEATs exhibiting stronger compositional representations than supervised or multimodal approaches. These results establish reference points for future work and highlight the role of pretraining objectives in shaping compositional structure.

Future work may investigate how compositionality metrics relate to downstream performance and extend this framework beyond synthetic data to natural recordings. Applying these ideas to multimodal audio–text or audio–visual representations is another promising direction.



\bibliographystyle{IEEEtran}
\bibliography{refs}

\end{document}